\documentclass[manuscript, trackchanges]{aastex62}

\usepackage[version=3]{mhchem}
\usepackage{chemfig}
\tabletypesize{\footnotesize}

\turnoffeditone
\turnoffedittwo

\received{}
\revised{}
\accepted{}

\shorttitle{}
\shortauthors{Iino et al.}

\begin{document}


\title{A belt-like distribution of gaseous hydrogen cyanide on Neptune's equatorial stratosphere detected by ALMA}


\author{Takahiro Iino}
\affil{Information Technology Center, The University of Tokyo, 2-11-16, Yayoi, Bunkyo, Tokyo 113-8658, Japan}
\email{iino@nagoya-u.jp}

\author{Hideo Sagawa}
\affil{Faculty of Science, Kyoto Sangyo University, Motoyama, Kamigamo, Kita-ku, Kyoto 603-8555, Japan}

\author{Takashi Tsukagoshi}
\affil{National Astronomical Observatory of Japan, 2-21-1 Osawa, Mitaka, Tokyo 181-8588, Japan}

\author{Satonori Nozawa}
\affil{ Institute for Space-Earth Environmental Research, Nagoya University,  Furo-cho, Chikusa-ku, Nagoya, Aichi 464-8601, Japan}



\begin{abstract}
We present a spatially resolved map of integrated- intensity and abundance of Neptune's stratospheric hydrogen cyanide (HCN). 
The analyzed data were obtained from the archived 2016 observation of the Atacama Large \edit1{Millimeter/submillimeter } Array.
A 0.\arcsec42 $\times$ 0.\arcsec39 synthesized beam, which is equivalent to a latitudinal resolution of  $\sim$20$\arcdeg$ at the disk center, was fine enough to resolve Neptune's 2.\arcsec24 diameter disk. 
After correcting the effect of different optical path lengths, a spatial distribution of HCN emissions is derived over Neptune's disk, and it clearly shows a band-like HCN enhancement at the equator. 
Radiative transfer analysis indicates that the HCN volume mixing ratio measured at the equator was 1.92 ppb above the 10$^{-3}$ bar pressure level, which is 40$\%$ higher than that measured at the southern middle and high latitudes. 
The spatial distribution of HCN can be interpreted as either the effect of the transportation of \ce{N2} from the troposphere by meridional atmospheric circulation, or an external supply such as cometary collisions (or both of these reasons).
From the meridional circulation point of view, the observed HCN enhancement on both the equator and the pole can be explained by the production and accumulation of HCN at the downward branches of the previously suggested two-cell meridional circulation models.
However, the HCN-depleted latitude of 60\arcdeg S does not match with the location of the upward branch of the two-cell circulation models. 

\end{abstract}

\keywords{planets and satellites: atmospheres --- submillimeter: planetary systems}

\section{Introduction}
The presence of hydrogen cyanide (HCN) characterizes Neptune's stratospheric composition.
The first detection of HCN was made by single-dish \edit1{millimeter/submillimeter } telescopes observing the rotational transitions of $J$=3--2 and 4--3 \citep{Marten1993,Rosenqvist1992}. 
The observed line widths of HCN were significantly narrow as $\sim$20 MHz, indicating that HCN is present in the upper stratosphere. 
The volume mixing ratio (VMR) determined from $J$=3--2 and 4--3 observations were 3.0$\times$10$^{-10}$ and 1.0$\times$10$^{-9}$, respectively. 
The HCN was restricted to be present at altitudes above the HCN condensation level, 3.5 mbar atmospheric pressure region. 
Subsequent sensitive disk-averaged observations also identified vertical distributions, and suggested that HCN was present in the region above 0.9--3 mbar, where the condensation cannot occur \citep{Marten2005, Rezac2014}. 

To explain the production of HCN in Neptune's atmosphere, \cite{Lellouch1994} developed a photo-chemical model of N-bearing species. 
The production of HCN from dissociated N-atoms was explained by two reactions: \ce{N + CH3 -> H2CN + H}, \ce{H2CN + H -> HCN + H2}. 
Two scenarios were proposed for the origin of the N-atoms: (1) infalling of ionized N-atoms transported from Neptune's largest moon, Triton, and (2) the upward transportation of \ce{N2} from the warm troposphere and subsequent dissociation to N-atoms by Galactic Cosmic Ray. 
The latter scenario suggests the importance of \ce{N2} transportation into the stratosphere by global circulation. 
The circulation has been inferred by the observations of continuum emissions and tropospheric gases because the atmospheric transportation may cause perturbations in the brightness temperature by adiabatic heating and cooling, and molecular opacity variation \citep{depater2014, Fletcher2014, Tollefson2019}.
Previous studies suggested that the observational signatures of the dry south polar troposphere, cold mid-latitudinal stratosphere and warm equatorial stratosphere are caused by the downward transportation of dry air, adiabatic cooling induced by the upward transportation, and adiabatic heating by the downward transportation, respectively. 
The suggested upward branch in the meridional circulation could transport \ce{N2} from the troposphere into the stratosphere in the mid-latitude, and possibly produce HCN by photo-chemical reactions in the stratosphere. 

In turn, some observations supported an external origin scenario that cometary impact and influx of the Interplanetary Dust Particle supply volatiles to the stratosphere. 
Such a process is well known for Jupiter, where the collisions of comet Shoemaker-Levy/9 produced a large amount of volatiles such as carbon monoxide (CO), HCN, carbon monosulfide (CS), and \ce{H2O}, as long-lived species in the stratosphere \citep{Lellouch1997, Moreno2001, Moreno2003, Cavalie2013, Iino2016c}.
Among these species, on Neptune, CS and \ce{H2O} are particularly important probes of the external volatiles supply because such species cannot pass through the cold tropopause in the gas phase (they are easily condensed in low-temperature environment). 
Some attempts have been made to detect CS on Neptune \citep{Moreno1998, Iino2014}, and a recent Atacama Large Millimeter/submillimeter Array (ALMA) observation reported the first detection of CS ($J$ = 7--6) on Neptune with a 2.4--21.0 $\times$ 10$^{-11}$ mixing ratio above the 0.5--0.03 mbar pressure level\citep{Moreno2017}. 
The detection of CS on Neptune reinforces the evidence for a previous cometary impact that should have supplied HCN, along with CS (and CO) at the same time as occurred on Jupiter.

HCN is also a subject of research for the Saturn's largest moon, Titan. HCN has been observed by in-situ, space- and ground-based observations. 
Those observations revealed a remarkable seasonal change in Titan's HCN distribution, in which the winter hemisphere has a larger abundance than that of the summer hemisphere \citep{Coustenis1989, Coustenis2005, Coustenis2010, Coustenis2016, Thelen2019}.
In particular, the Cassini spacecraft illustrated HCN enhancement at the winter pole before the summer solstice. 
HCN abundance measured at 75\arcdeg S showed $\sim$1000 times enhancement in two years, from 2012 to 2014 \citep{Coustenis2016}. 
The inhomogeneous distribution is attributed to effects of global circulation and photo-chemistry.
\edit1{ \cite{Vinatier2015} successfully obtained the seasonal evolution of HCN vertical distribution by analyzing the number of Cassini observation data using the radiative transfer method. }
They concluded that, in 2011, two years after the vernal equinox, a single north-to-south-pole cell appeared in the meridional circulation, and transported HCN-enriched air to the south pole. 

The above-described case of Titan is indicative that the spatially resolved observation of volatiles, in particular HCN, could give us a new clue on the atmospheric dynamics and chemistry of Neptune. 
The ALMA achieves high spatial resolution observation of solar system objects in millimeter and submillimeter wavelength, whereas previous observations using single-dish telescopes were able to obtain only the disk-averaged spectra of HCN. 
In this paper, we first report the spatial distribution of HCN in Neptune's stratosphere obtained with ALMA. 

\section{Image synthesis of ALMA archived data}
We analyzed an archived ALMA data of project ID 2015.1.01471.S (PI: R. Moreno) including the HCN ($J$=4--3) rotational transition at 354.505 GHz, the same project that was used in \cite{Moreno2017}.
The observations were performed originally to search for isotopologues of major species such as CO and HCN, and minor chemical species such as CS and \ce{CH3CCH} on Neptune.  

The observation was performed on 30 April 2016 (UTC) using 41 12--m antennae. 
At the observed time, the apparent angular diameter of Neptune was 2.$\arcsec$24.
Both the sub-observer and sub-solar latitude were 26$\arcdeg$S. 
The center frequency of the spectral window used for the analysis was 355.19004 GHz. 
The total bandwidth of the spectral window and the channel spacing were 1875 and 0.977 MHz (the effective spectral resolution was $\sim$1.13 MHz), respectively. 
The total observing time of Neptune was 808.1 seconds. 

The Common Astronomy Software Applications package, version 4.6, was used for the reduction of u-v data.
We calibrated the raw data using the scripts provided by the ALMA.
The flux, bandpass and phase calibrators were the asteroid Pallas, \edit2{quasars} J0006-0623 and J2246-1206, respectively.  
In the u-v dataset, the continuum and line emissions were separated.
A CLEAN procedure was used for imaging with the following parameters: 320$\times$320 pixels with 0.$\arcsec$025 pixel spacing, "natural" weighting, 0.1 mJy threshold, 0.1 gain and csclean mode. 
A circular CLEAN region that had a diameter slightly larger than Neptune was employed with the channel (width = 1) clean mode. 
The achieved synthesized beam size was 0.$\arcsec$42$\times$0.$\arcsec$39, which was fine enough to resolve the Neptune's disk spatially. 

\section{Analysis of spatial distribution of HCN emission}
\begin{figure*}  
 \begin{center}
  \includegraphics[width=8cm]{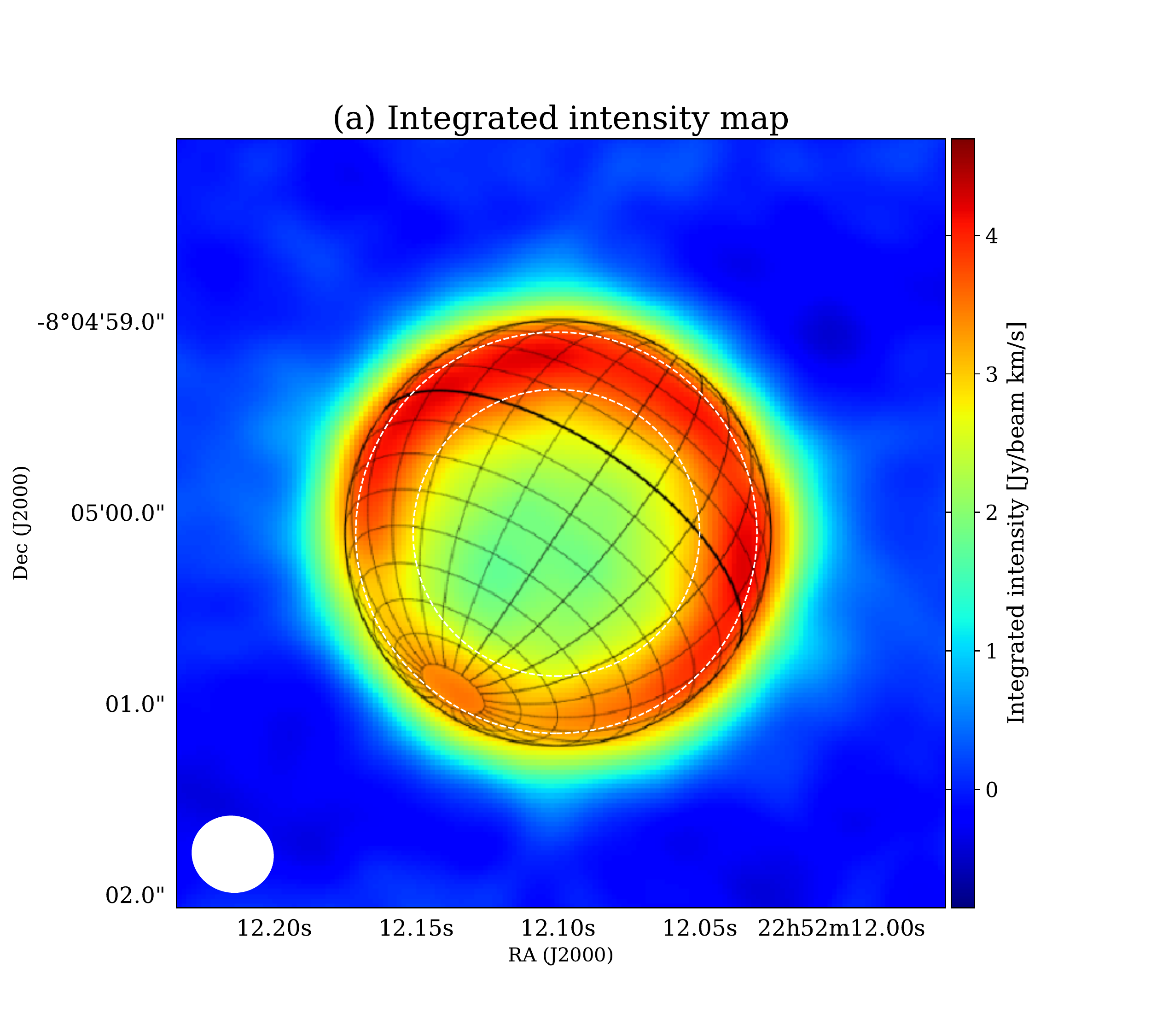}
    \includegraphics[width=8cm]{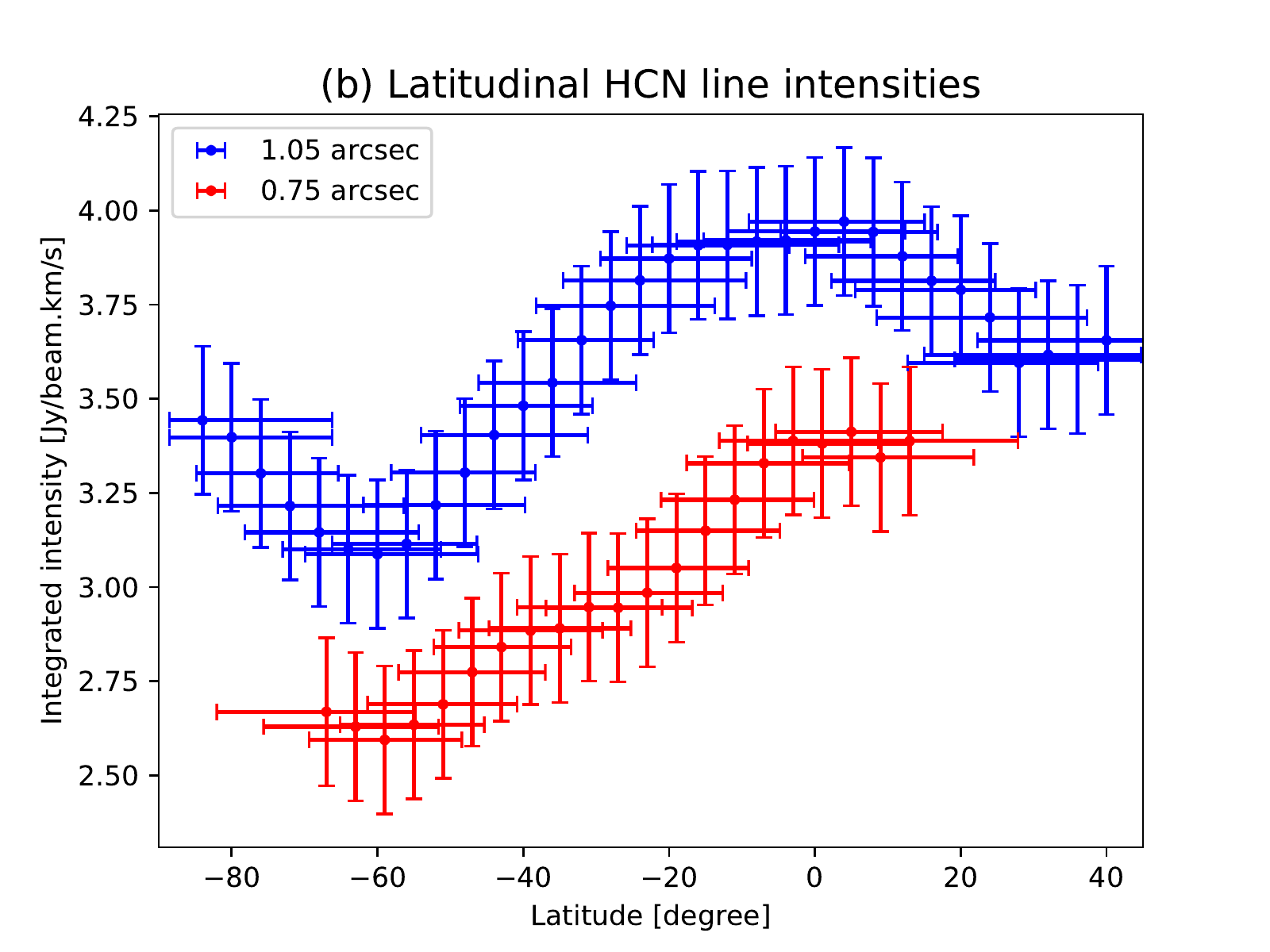}
   \includegraphics[width=8cm]{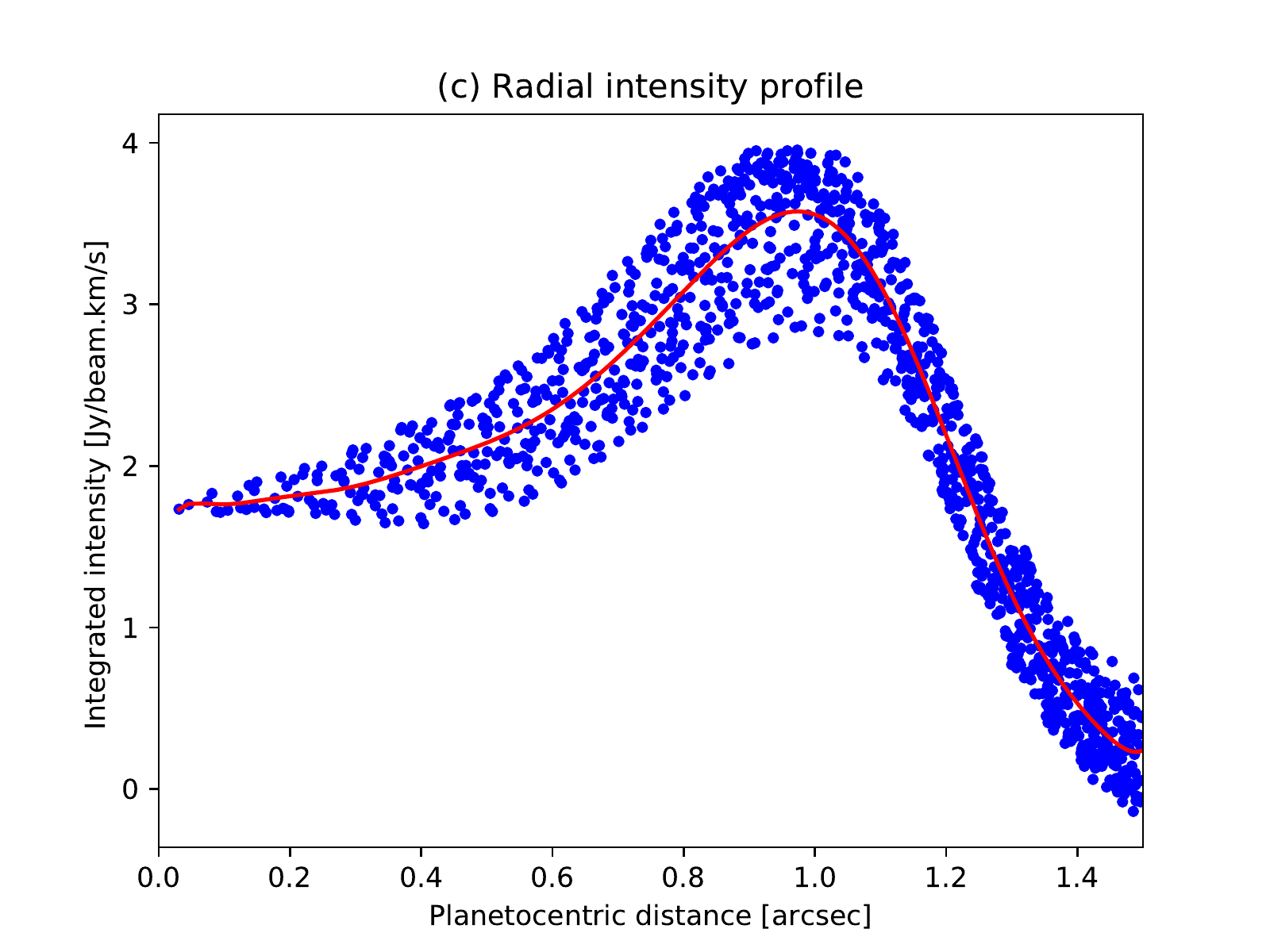}
 \end{center}
 \caption{(a) Integrated intensity map of HCN ($J$=4--3) for Neptune. \edit1{The white ellipse at the bottom left illustrates the shape of the synthesized beam.}   (b) HCN intensity profiles along the 1\arcsec.05 and 0 \arcsec.75 radius circle indicated by white dashed lines in (a). (c) Radial HCN intensity profile (blue dots) and averaged profile (red curve) . Horizontal and vertical error bars are the same as the synthesized beam size, $\sim$0\arcsec.2 and 0.2 \edit1{Jy/beam km/s}, respectively.}
 \label{mom0}
\end{figure*}

To illustrate the spatial distribution of HCN on Neptune's disk, an integrated-intensity map that integrates the $\pm$30 MHz frequency range, which covers the entire HCN emission line, was produced. 
The map is shown in Figure \ref{mom0}(a) and exhibits a clear ring-like structure with a $\sim$0.\arcsec95 radius, which has been also reported in an unpublished work using the Sub Millimeter Array \citep{Moullet2011}.
Note that the ring structure is attributed to the increase of the line-of-sight path length in the Neptune atmosphere as the emission angle increases.
Figure \ref{mom0}(b) shows the HCN intensity measured at the same emission angle along the 1.\arcsec05-- and 0.\arcsec75-- radius circle in Figure \ref{mom0}(a), which can exhibit latitudinal intensity variation without variation of the emission angle. 
Vertical and horizontal error bars are r.m.s. noise level measured outside the disk and the latitude range included in the synthesized beam, respectively.
Eastern and western hemispheric intensities are averaged. 
For both selected circles, an interesting feature is that the greatest intensity peak is locating at the equator. 
The lowest intensity values of the both circles locate at 60\arcdeg S.
The peak intensity measured at the equator is 25--30$\%$ higher than the lowest value. 
In addition, for the 1.\arcsec05 circle, a weak intensity peak is also found on the south pole. 
The intensity difference between 60\arcdeg S and the south pole is $\sim$10$\%$. 

\begin{figure*}  
 \begin{center}
  \includegraphics[width=8cm]{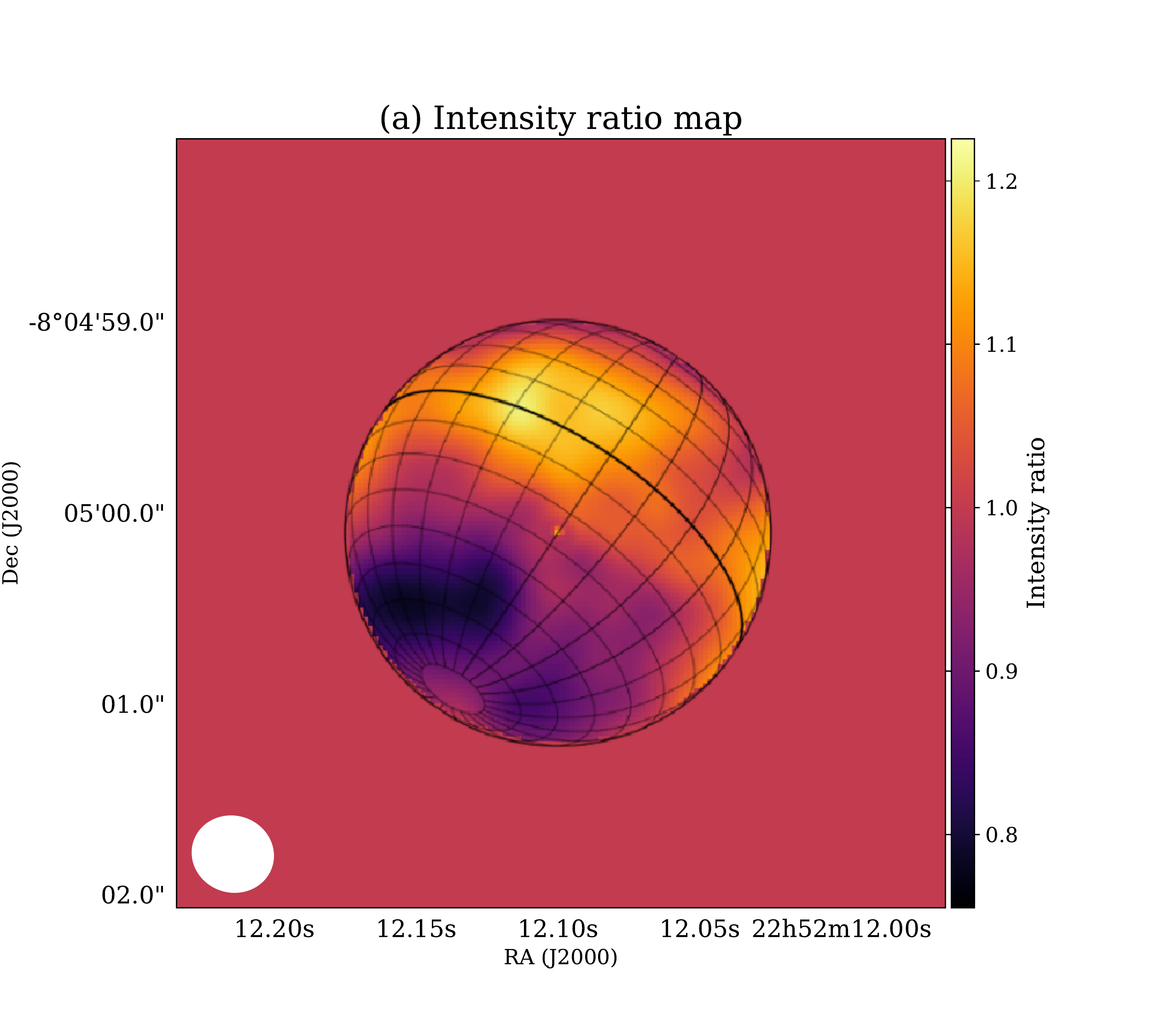}
  \includegraphics[width=8cm]{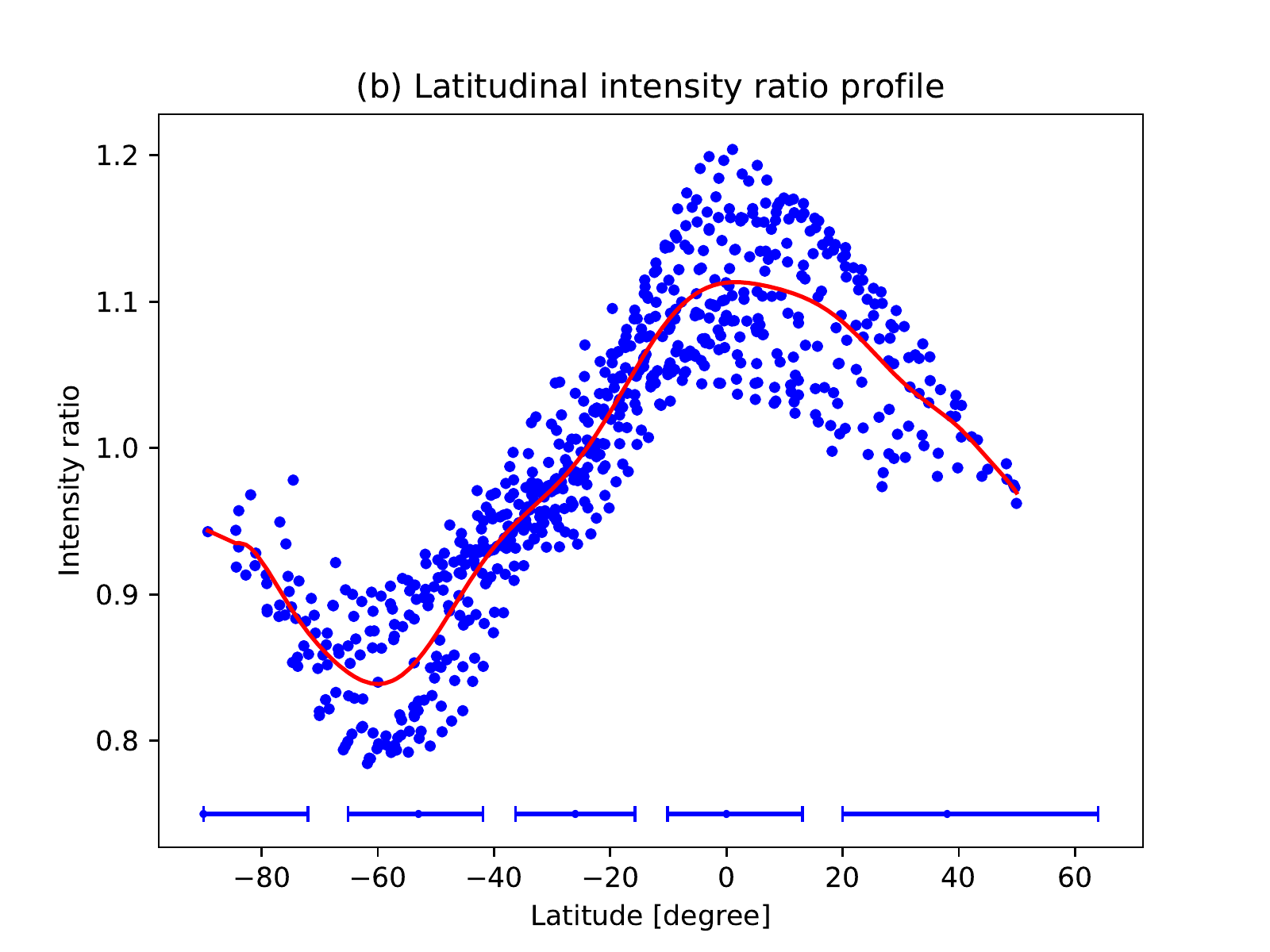}
 \end{center}
 \caption{(a) Intensity ratio (measured value versus radially averaged profile) map of HCN. \edit1{The white ellipse at the bottom left illustrates the shape of the synthesized beam.} (b) Latitudinal profile of the intensity ratio. Blue dots and red curve are for each pixel and the latitudinally averaged profile, respectively. Horizontal error bars at the bottom are latitudinal error measured on the central meridian line. }
 \label{radial_profile}
\end{figure*}

Figure \ref{mom0}(c) shows the radial HCN intensity profile measured for each pixel and radially averaged profile. 
A peak located at $\sim$1$\arcsec$ corresponds to the ring structure shown in Figure  \ref{mom0}(a).
The radially averaged profile was produced by the polynomial fitting method. 
The intensity ratio of measured and averaged intensity is shown in Figure \ref{radial_profile}. 
While both the vertical and horizontal errors are omitted due to the dense distribution of dots, corresponding errors are same as in \edit1{Figure \ref{mom0}(b) } and the size of the synthesized beam size, respectively. 

To illustrate the latitudinal intensity distribution over the entire disk, an intensity ratio map of measured intensity versus radially averaged intensity was produced (Figure \ref{radial_profile} (a)).
The same method was applied to derive the global continuum emission \edit1{distribution } on Neptune \citep{Iino2018a}. 
On the equator, a belt-like HCN-rich region, which corresponds to the latitudinal peaks found in the 1\arcsec.05 radius circle in Figure \ref{mom0} (b), is clearly shown. 
At the southern mid-latitude of $\sim$60\arcdeg S, a low intensity ratio region is found in the western hemisphere. 
A latitudinal profile of the intensity ratio is shown in Figure \ref{radial_profile} (b). 
For a better visibility of the figure, \edit1{every third pixel intensity was plotted.}
The derived structure is similar to that measured along the same emission angle as shown in Figure \ref{mom0} (b). 
The red curve represents the latitudinally averaged profile of the ratio. 
The equatorial peak shows a latitudinally symmetrical structure. 
In addition, a relatively weak peak is found at the south pole. 
Latitudinal errors corresponding to the synthesized beam size are represented at the bottom of Figure \ref{radial_profile} (b). 
At the 60\arcdeg S region, intensity ratio values can be divided into two groups, which are likely to correspond to a dark spot located in the western hemisphere and other regions on 60\arcdeg S arc. 
Considering the self rotation period of Neptune of $\sim$16 hours, Neptune rotates $\sim$15$\arcdeg$ during the observation time. 
In turn, difference of intensity between bright and dark region on 60\arcdeg S arc is $\sim$10$\%$, which is similar to the systematic error value of ALMA's intensity measurement accuracy. 
Thus, detection of the HCN depletion spot on 60\arcdeg S arc is marginal. 

\section{Radiative transfer analysis}

The radiative transfer method was employed to estimate the latitudinal difference of HCN abundance between 60\arcdeg S and the equatorial region by searching the best-fit spectrum.
For the calculation, we employed the open-source software Planetary Spectrum Generator (PSG) \citep{Villanueva2018}. 
Because PSG has an online \edit1{Application Program Interface } that is easy to use, one can evaluate our result by reproducing the observed spectra. 

Pressure levels of modeled vertical atmospheric structure were ranged from 100 to 10$^{-4}$ bar with 40 layers. 
For the temperature profile, we employed a disk-averaged result retrieved from \citep{Fletcher2010}.
Gaseous \ce{H2}, He, \ce{CH4}, CO and HCN were considered as the atmospheric constituents. 
\edit1{Their vertical abundance profile except for HCN were the initial set of PSG. }
\edit2{\citep{Moses2005, Marten2005, Lellouch2010}}
Spectroscopic parameters are as of HITRAN database. 
Because PSG can employ a horizontally symmetrical beam, a 0.\arcsec41 diameter beam was employed while the true beam shape is slightly elliptical.
We attempted to obtain the HCN abundance at two points on 60\arcdeg S and the equator. 

As the vertical HCN profile, a constant molecular VMR above a specific pressure level, $p_0$, was employed in this study.
\edit1{We have tested some $p_0$ values in the range from 0.5 to 2.0 mbar pressure level.
As a result, }we employed 1.0 mbar pressure level as $p_0$ because the value could reproduce the observed spectra to a relatively better extent.  

The best-fit spectrum for each point was identified within the VMR parameter space by the least-square method.
The $\Delta\chi^2$ technique was used to obtain the error for the fitted parameter \edit1{\citep{Teanby2006}}. 
The 1--$\sigma$ significance level was used as the error value, corresponding to $\Delta\chi^2$ = 1.0.
The derived VMR and errors for 60\arcdeg S and the equatorial region were 1.17$\pm$0.03 and 1.66$^{+0.06}_{-0.03}$ ppb, respectively, above the 1.0 mbar pressure region.
Thus, the equatorial region is determined to have 40$\%$ higher abundance than that of 60\arcdeg S region. 

\begin{figure*}  
 \begin{center}
\includegraphics[width=8cm]{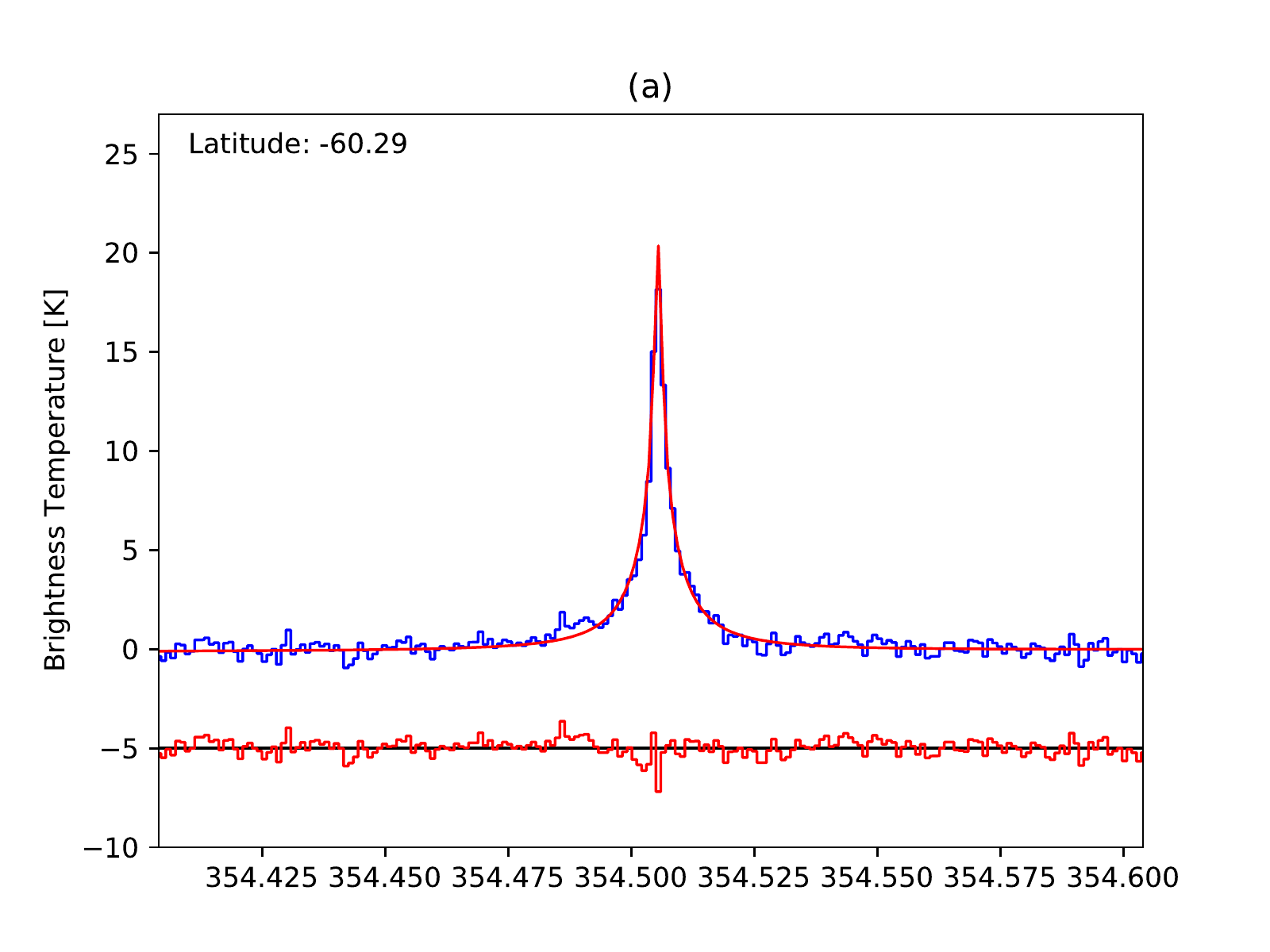}
\includegraphics[width=8cm]{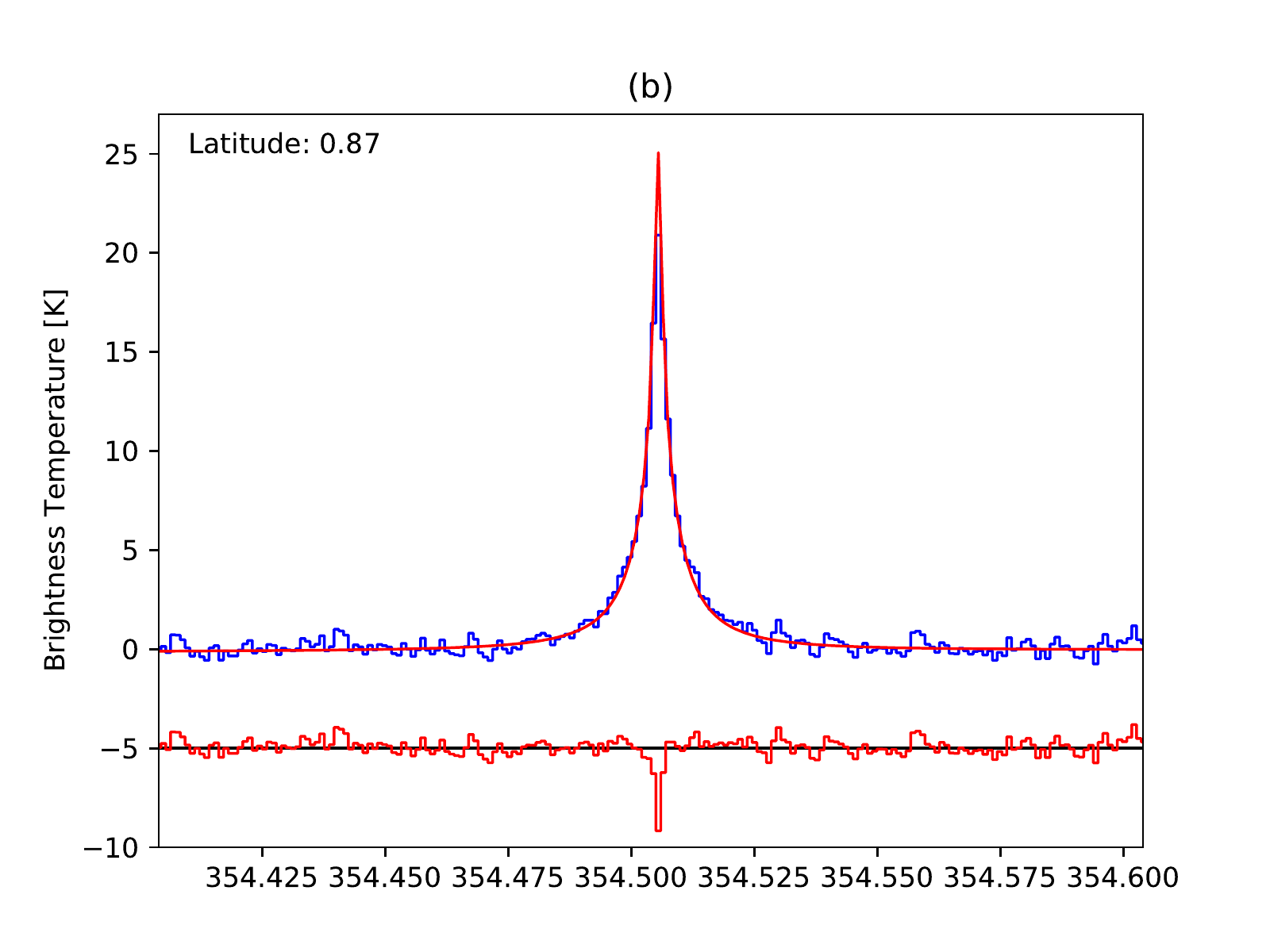}
 \end{center}
 \caption{Observed (blue) and modeled (red) HCN spectra measured and derived at 60\arcdeg S(a) and the equator(b) on the central meridian line.  Residual is plotted with separation of -5 K in red curve. }
 \label{spectra}
\end{figure*}

Figure \ref{spectra}(a) and (b) shows a series of the measured and best-fit spectra.
Residuals between the observed and modeled spectra are also shown as red curves with an offset of -5 K.
\edit1{Considering the atmospheric structure, opacity of the HCN line core is derived to be $\sim$0.3. 
Because of the thin opacity, the line core is likely to have the sensitivity peak nearly at 1.0 mbar level. }
For the equatorial spectra, the $\pm$1.5 MHz frequency region from the line center could not fit well.
This result leads to three possible scenarios, that the HCN abundance or temperature decreases, or both decrease in high altitude region.
However, in this study, we did not attempt to obtain more specific abundance or temperature profile, because the area of the line center residual is quite small compared with the total area of the emission spectra and the possible effect on the retrieved column density is very limited.

\section{Discussion}
This new analysis of the spatially-resolved ALMA spectroscopic observation data for Neptune indicates that Neptune's stratospheric HCN has a horizontally non-uniform distribution, in which the equator has a $\sim$30$\%$ higher line intensity and $\sim$40$\%$ higher abundance than that of the 60\arcdeg S region.
Possible scenarios that might explain the observed results are discussed from two point of views of the origin of HCN: internal and external sources.

\subsection{Possible sources of HCN intensity gradient}
Spatial variations in HCN line intensity can be caused both by spatial variations in the temperature of the foreground stratosphere and the background troposphere, and differences in HCN abundance.
The variation in stratospheric or tropospheric temperature is equivalent to that of the HCN intensity because the HCN line \edit1{core opacity} of $\sim$0.3 is optically thin; thus, its intensity is roughly linear with the temperature variation.
To explain the observed maximum 30$\%$ intensity variation only by the temperature variation, the spatial temperature difference at the same altitude should have nearly the same ratio as the integrated line intensity variation. 
\cite{Fletcher2014} retrieved atmospheric temperatures from two observations of Voyager 2 in 1989 and Keck/\edit1{Long Wavelength Spectrometer} in 2003. 
Above 1.0 mbar, where HCN molecules are present, the inferred temperature variations were no more than 5 K, which are not enough to explain the observed HCN intensity variation. 
Note that the tropospheric temperature variation \edit1{measured by 245 and 646 GHz ALMA spatially-resolved continuum observation} were reported to be smaller than that expected in the stratosphere \citep{Tollefson2019,Iino2018a}. 
Thus, in this study, we consider that the variation in HCN line intensity was caused mainly by a variation in HCN abundance.

\subsection{Implication for global circulation}
The spatial distribution of short-lived trace species is a powerful tool for investigating atmospheric dynamics.
The obtained HCN latitudinal gradient shows a morphological similarity with that observed on Titan, where the global single-cell circulation possibly produces a latitudinal abundance difference in which the winter and summer hemispheres exhibit the highest and lowest peaks of HCN, respectively  \citep{Coustenis1989, Coustenis2005, Coustenis2010, Coustenis2016, Thelen2019}.
On Titan, similar to the terrestrial  Brewer--Dobson circulation, a global single cell in the meridional circulation is likely to transport N-bearing species depleted air parcel from summer to winter hemisphere. 
Various N-bearing species are being produced during the horizontal transportation, and accumulated on the winter pole where the subsiding transportation is present. 
In addition, the result obtained here is similar to that for Earth's stratospheric ozone (e.g. \cite{Wayne2000}), which is also caused by a summer to \edit1{winter } single cell meridional circulation. 

On Neptune, various observation techniques have already been used to propose the presence of a global tropospheric and stratospheric circulation. 
A re-analysis of Voyager/\edit1{Infrared Interferometer Spectrometer and Radiometer } spectra,  \cite{Fletcher2014} concluded that a global circulation that cold air rises in the mid-latitude and subsides both on the equator and the poles. 
A warm tropospheric equator and pole are likely to be produced by the adiabatic heating induced by the subsiding air. 
Similarly, \cite{depater2014} suggested that upward transportation at mid-latitude, $\sim$40\arcdeg S,  creates a belt-like structure of the tropospheric cloud that is caused by the adiabatic cooling of the rising air parcel. 
A recent ALMA continuum observation expects that mid-latitude upward transportation is present at relatively lower latitude region, $\sim$30\arcdeg S \citep{Tollefson2019}.
From a morphological point of view, our obtained HCN distribution map can be connected to the previous observations of the global circulation as follows: the high abundances at the equator and south pole are likely to be due to the accumulation of HCN produced during the horizontal transportation in the same manner as the mechanism of Titan's HCN distribution.
Considering the troposphere--stratosphere circulation, \ce{N2} transported to the stratosphere dissociates into N-atom by the photolysis, and leads to HCN production via \ce{H2CN}, as mentioned in the Introduction section. 
HCN enhancements at both the equator and south pole are quite consistent with a two-cell circulation model for the southern hemisphere.

In turn, the HCN-depleted region observed at 60\arcdeg S disagrees with the location of the upward branch of the two-cell circulation model (the circulation model shows the upward transportation at 40\arcdeg S).
Although the reason for this remains for further study, it should be noted that our study probes the upper stratosphere while the previously suggested two-cell meridional circulation model has been mainly based on observations of the upper troposphere. An effective use of HCN maps may bring an additional constraint to the atmospheric circulation at higher altitudes.
\edit1{In addition, because of the coarse resolution of this analysis, 40\arcdeg S and 60\arcdeg S region were not resolved clearly. Thus, a new observation with a higher spatial resolution enables us to determine the latitude of the HCN-depleted belt. }

\subsection{External source model}
A large cometary impact is also a possible cause of the observed HCN distribution.
Strong evidence for such an impact was previously provided by the presence of CS \citep{Moreno2017} and by the CO-rich upper stratosphere \citep{Lellouch2005, Hesman2007, Fletcher2010}.
Because no S-bearing species can be supplied from the troposphere, CS is considered to have only a cometary origin. 
Thus, it is also possible for HCN to be supplied by a past cometary impact; likely an impact at the equator.
To evaluate the cometary impact hypothesis, a new analysis of the latitudinal distribution of CS is crucial. 
If CS shows the equatorial enhancement seen in HCN, the impact hypothesis is strongly supported. 
It is noted that our observation shows HCN enhancement at the south pole as well. 
Such a latitudinal distribution does not fit well with a simple meridional diffusion of HCN from a single collision.

\subsection{Future perspectives}
This study presented that a spatially resolved HCN observation has significant potential for providing various information for discerning the atmospheric circulation and/or a past cometary impact.
As mentioned in Section 5.3, highly sensitive observation of CS to illustrate its spatial distribution, along with the determination of the 3-D distribution of CO, would strongly constrain the cometary impact scenario. 
Also the spatial distribution of \ce{H2O}, which is not observable by ALMA, is crucial to evaluate the scenario as for the case for Jupiter \citep{Cavalie2013}.
In addition, precise observations to determine the $^{14}$N/$^{15}$N isotopic ratio in HCN may constrain its source and production process. 
For example, on Jupiter, a large nitrogen isotopic fractionation in HCN (4.3-16.7 times higher than the typical solar system value) was detected \citep{Matthews2002}. 
This fractionation may be caused by the thermo-chemical processed induced by the cometary collision.
In addition, on Titan, a different isotopic fractionation between \ce{N2} and its daughter species, HCN, is known \citep{Hidayat1997, Gurwell2004, Molter2016}. 
A theoretical chemical model suggests that the nitrogen-bearing species is fractionated in a different extent according to different dissociation processes of \ce{N2} \citep{Dobrijevic2018}. 
These applications to other planets and satellites lead one to expect that a new determination of the nitrogen isotopic ratio in Neptune HCN will provide new implication on its origin.

\acknowledgments
This paper makes use of the following ALMA data: ADS/JAO.ALMA$\#$2015.1.01471.S. 
TI wishes to thank Satoru Nakamoto and Yuma Nakayama of Nagoya University for their dedicated contribution on the preliminary analysis of the ALMA data. 

This work was financially supported by a Telecommunications Advancement Foundation, JSPS Kakenhi (17K14420 and 19K14782) and an Astrobiology Center Program of National Institutes of Natural Sciences grants.
ALMA is a partnership of ESO (representing its member states), NSF (USA) and NINS (Japan), together with NRC (Canada), MOST and ASIAA (Taiwan), and KASI (Republic of Korea), in cooperation with the Republic of Chile. The Joint ALMA Observatory is operated by ESO, AUI/NRAO and NAOJ.

\software{Astropy \citep{Robitaille2013}}
\software{PSG \citep{Villanueva2018}}

\clearpage

\clearpage

\bibliography{library}

\begin{thebibliography}{}
\expandafter\ifx\csname natexlab\endcsname\relax\def\natexlab#1{#1}\fi
\providecommand{\url}[1]{\href{#1}{#1}}

\bibitem[{Cavali{\'{e}} {et~al.}(2013)Cavali{\'{e}}, Feuchtgruber, Lellouch,
  de~Val-Borro, Jarchow, Moreno, Hartogh, Orton, Greathouse, Billebaud,
  Dobrijevic, Lara, Gonz{\'{a}}lez, \& Sagawa}]{Cavalie2013}
Cavali{\'{e}}, T., Feuchtgruber, H., Lellouch, E., {et~al.} 2013, Astronomy
  {\&} Astrophysics, 553, A21.
\newblock \url{http://www.aanda.org/10.1051/0004-6361/201220797}

\bibitem[{Coustenis {et~al.}(1989)Coustenis, B??zard, \&
  Gautier}]{Coustenis1989}
Coustenis, A., B??zard, B., \& Gautier, D. 1989, Icarus, 82, 67

\bibitem[{Coustenis {et~al.}(2005)Coustenis, Irwin, Teanby, Jennings, Romani,
  Be, Vinatier, Strobel, Calcutt, Read, Fouchet, Gautier, Lellouch, Marten, \&
  Prange}]{Coustenis2005}
Coustenis, a., Irwin, P. G.~J., Teanby, N.~a., {et~al.} 2005, 975

\bibitem[{Coustenis {et~al.}(2010)Coustenis, Jennings, Nixon, Achterberg,
  Lavvas, Vinatier, Teanby, Bjoraker, Carlson, Piani, Bampasidis, Flasar, \&
  Romani}]{Coustenis2010}
Coustenis, a., Jennings, D.~E., Nixon, C.~a., {et~al.} 2010, Icarus, 207, 461.
\newblock \url{http://dx.doi.org/10.1016/j.icarus.2009.11.027}

\bibitem[{Coustenis {et~al.}(2016)Coustenis, Jennings, Achterberg, Bampasidis,
  Lavvas, Nixon, Teanby, Anderson, Cottini, \& Flasar}]{Coustenis2016}
Coustenis, A., Jennings, D.~E., Achterberg, R.~K., {et~al.} 2016, Icarus, 270,
  409.
\newblock \url{http://dx.doi.org/10.1016/j.icarus.2015.08.027}

\bibitem[{de~Pater {et~al.}(2014)de~Pater, Fletcher, Luszcz-Cook, DeBoer,
  Butler, Hammel, Sitko, Orton, \& Marcus}]{depater2014}
de~Pater, I., Fletcher, L.~N., Luszcz-Cook, S., {et~al.} 2014, Icarus, 237,
  211.
\newblock \url{http://dx.doi.org/10.1016/j.icarus.2014.02.030}

\bibitem[{Dobrijevic \& Loison(2018)}]{Dobrijevic2018}
Dobrijevic, M., \& Loison, J.~C. 2018, Icarus, 307, 371.
\newblock \url{https://doi.org/10.1016/j.icarus.2017.10.027}

\bibitem[{Fletcher {et~al.}(2014)Fletcher, de~Pater, Orton, Hammel, Sitko, \&
  Irwin}]{Fletcher2014}
Fletcher, L.~N., de~Pater, I., Orton, G.~S., {et~al.} 2014, Icarus, 231, 146.
\newblock \url{http://linkinghub.elsevier.com/retrieve/pii/S0019103513005095}

\bibitem[{Fletcher {et~al.}(2010)Fletcher, Drossart, Burgdorf, Orton, \&
  Encrenaz}]{Fletcher2010}
Fletcher, L.~N., Drossart, P., Burgdorf, M., Orton, G.~S., \& Encrenaz, T.
  2010, Astronomy and Astrophysics, 514, A17.
\newblock \url{http://www.aanda.org/10.1051/0004-6361/200913358}

\bibitem[{Gurwell(2004)}]{Gurwell2004}
Gurwell, M.~a. 2004, The Astrophysical Journal, 616, L7

\bibitem[{Hesman {et~al.}(2007)Hesman, Davis, Matthews, \& Orton}]{Hesman2007}
Hesman, B.~E., Davis, G.~R., Matthews, H.~E., \& Orton, G.~S. 2007, Icarus,
  186, 342.
\newblock \url{http://linkinghub.elsevier.com/retrieve/pii/S0019103506003149}

\bibitem[{Hidayat {et~al.}(1997)Hidayat, Marten, B{\'{e}}zard, \&
  Gautier}]{Hidayat1997}
Hidayat, T., Marten, a., B{\'{e}}zard, B., \& Gautier, D. 1997, Icarus, 182,
  170.
\newblock
  \url{http://www.sciencedirect.com/science/article/pii/S0019103596956407}

\bibitem[{Iino {et~al.}(2014)Iino, Mizuno, Nakajima, Hidemori, Tsukagoshi, \&
  Kato}]{Iino2014}
Iino, T., Mizuno, A., Nakajima, T., {et~al.} 2014, Planetary and Space Science,
  104, doi:10.1016/j.pss.2014.09.013

\bibitem[{Iino {et~al.}(2016)Iino, Ohyama, Hirahara, Takahashi, \&
  Tsukagoshi}]{Iino2016c}
Iino, T., Ohyama, H., Hirahara, Y., Takahashi, T., \& Tsukagoshi, T. 2016, The
  Astronomical Journal, 152, 179.
\newblock
  \url{http://stacks.iop.org/1538-3881/152/i=6/a=179?key=crossref.a0f228dfac9f4b6e423c2b067a62f7a9}

\bibitem[{Iino \& Yamada(2018)}]{Iino2018a}
Iino, T., \& Yamada, T. 2018, The Astronomical Journal, 155, 92.
\newblock
  \url{http://stacks.iop.org/1538-3881/155/i=2/a=92?key=crossref.aec8f8bea3e7e79480aa1846181fecce}

\bibitem[{Lellouch(1994)}]{Lellouch1994}
Lellouch, E. 1994, Icarus, 108, 112.
\newblock
  \url{http://linkinghub.elsevier.com/retrieve/doi/10.1006/icar.1994.1045}

\bibitem[{Lellouch {et~al.}(2005)Lellouch, Moreno, \& Paubert}]{Lellouch2005}
Lellouch, E., Moreno, R., \& Paubert, G. 2005, Astronomy and Astrophysics, 40,
  37

\bibitem[{Lellouch {et~al.}(1997)Lellouch, B{\'{e}}zard, Moreno,
  Bockel{\'{e}}e-Morvan, Colom, Crovisier, Festou, Gautier, Marten, \&
  Paubert}]{Lellouch1997}
Lellouch, E., B{\'{e}}zard, B., Moreno, R., {et~al.} 1997, Planetary and Space
  Science, 45, 1203.
\newblock \url{http://linkinghub.elsevier.com/retrieve/pii/S0032063397000433}

\bibitem[{Lellouch {et~al.}(2010)Lellouch, Hartogh, Feuchtgruber,
  Vandenbussche, de~Graauw, Moreno, Jarchow, Cavali{\'{e}}, Orton,
  Banaszkiewicz, Blecka, Bockel{\'{e}}e-Morvan, Crovisier, Encrenaz, Fulton,
  K{\"{u}}ppers, Lara, Lis, Medvedev, Rengel, Sagawa, Swinyard, Szutowicz,
  Bensch, Bergin, Billebaud, Biver, Blake, Blommaert, Cernicharo, Courtin,
  Davis, Decin, Encrenaz, Gonzalez, Jehin, Kidger, Naylor, Portyankina,
  Schieder, Sidher, Thomas, de~Val-Borro, Verdugo, Waelkens, Walker, Aarts,
  Comito, Kawamura, Maestrini, Peacocke, Teipen, Tils, \&
  Wildeman}]{Lellouch2010}
Lellouch, E., Hartogh, P., Feuchtgruber, H., {et~al.} 2010, Astronomy and
  Astrophysics, 518, L152.
\newblock \url{http://www.aanda.org/10.1051/0004-6361/201014600}

\bibitem[{Marten {et~al.}(1993)Marten, Gautier, Owen, Sanders, Matthews,
  Atreya, Tilanus, \& Deane}]{Marten1993}
Marten, A., Gautier, D., Owen, T., {et~al.} 1993, The Astrophysical Journal,
  406, 285.
\newblock \url{http://adsabs.harvard.edu/doi/10.1086/172440}

\bibitem[{Marten {et~al.}(2005)Marten, Matthews, Owen, Moreno, Hidayat, \&
  Biraud}]{Marten2005}
Marten, A., Matthews, H., Owen, T., {et~al.} 2005, Astronomy and Astrophysics,
  429, 1097

\bibitem[{Matthews {et~al.}(2002)Matthews, Marten, Moreno, \&
  Owen}]{Matthews2002}
Matthews, H.~E., Marten, A., Moreno, R., \& Owen, T. 2002, The Astrophysical
  Journal, 580, 598.
\newblock \url{http://stacks.iop.org/0004-637X/580/i=1/a=598
  https://iopscience.iop.org/article/10.1086/343108}

\bibitem[{Molter {et~al.}(2016)Molter, Nixon, Cordiner, Serigano, Irwin,
  Teanby, Charnley, \& Lindberg}]{Molter2016}
Molter, E.~M., Nixon, C.~A., Cordiner, M.~A., {et~al.} 2016, The Astronomical
  Journal, 152, 42.
\newblock
  \url{http://stacks.iop.org/1538-3881/152/i=2/a=42?key=crossref.4259a4ec01fd61a24d2bd0bf6ad0d077}

\bibitem[{Moreno(1998)}]{Moreno1998}
Moreno, R. 1998, PhD thesis

\bibitem[{Moreno {et~al.}(2017)Moreno, Lellouch, Cavali{\'{e}}, \&
  Moullet}]{Moreno2017}
Moreno, R., Lellouch, E., Cavali{\'{e}}, T., \& Moullet, A. 2017, Astronomy
  {\&} Astrophysics, 608, L5.
\newblock \url{http://www.aanda.org/10.1051/0004-6361/201731472
  http://www.aanda.org/10.1051/0004-6361:20052990{\%}0Ahttp://www.aanda.org/10.1051/0004-6361/201731472}

\bibitem[{Moreno {et~al.}(2001)Moreno, Marten, Biraud, B{\'{e}}zard, Lellouch,
  Paubert, \& Wild}]{Moreno2001}
Moreno, R., Marten, A., Biraud, Y., {et~al.} 2001, Planetary and Space Science,
  49, 473.
\newblock \url{http://linkinghub.elsevier.com/retrieve/pii/S0032063300001392}

\bibitem[{Moreno {et~al.}(2003)Moreno, Marten, Matthews, \&
  Biraud}]{Moreno2003}
Moreno, R., Marten, A., Matthews, H.~E., \& Biraud, Y. 2003, Planetary and
  Space Science, 51, 591.
\newblock \url{http://linkinghub.elsevier.com/retrieve/pii/S0032063303000722}

\bibitem[{Moses(2005)}]{Moses2005}
Moses, J.~I. 2005, Journal of Geophysical Research, 110, 1.
\newblock \url{http://www.agu.org/pubs/crossref/2005/2005JE002411.shtml}

\bibitem[{Moullet \& Gurwell(2011)}]{Moullet2011}
Moullet, A., \& Gurwell, M. 2011, EPSC-DPS Joint Meeting.
\newblock \url{http://yly-mac.gps.caltech.edu/A{\_}DPS/dps 2011 /a{\_}dps 2011
  program + abstracts/pdf/EPSC-DPS2011-1153-3.pdf
  http://yly-mac.gps.caltech.edu/A{\_}DPS/dps 2011 /a{\_}dps program +
  abstracts/pdf/EPSC-DPS2011-1153-3.pdf}

\bibitem[{Rezac {et~al.}(2014)Rezac, de~Val-Borro, Hartogh, Cavali{\'{e}},
  Jarchow, Rengel, \& Dobrijevic}]{Rezac2014}
Rezac, L., de~Val-Borro, M., Hartogh, P., {et~al.} 2014, Astronomy {\&}
  Astrophysics, 563, A4.
\newblock \url{http://www.aanda.org/10.1051/0004-6361/201323300}

\bibitem[{Robitaille {et~al.}(2013)Robitaille, Tollerud, Greenfield,
  Droettboom, Bray, Aldcroft, Davis, Ginsburg, Price-Whelan, Kerzendorf,
  Conley, Crighton, Barbary, Muna, Ferguson, Grollier, Parikh, Nair,
  G{\"{u}}nther, Deil, Woillez, Conseil, Kramer, Turner, Singer, Fox, Weaver,
  Zabalza, Edwards, {Azalee Bostroem}, Burke, Casey, Crawford, Dencheva, Ely,
  Jenness, Labrie, Lim, Pierfederici, Pontzen, Ptak, Refsdal, Servillat, \&
  Streicher}]{Robitaille2013}
Robitaille, T.~P., Tollerud, E.~J., Greenfield, P., {et~al.} 2013, Astronomy
  {\&} Astrophysics, 558, A33.
\newblock \url{http://www.aanda.org/10.1051/0004-6361/201322068}

\bibitem[{Rosenqvist {et~al.}(1992)Rosenqvist, Lellouch, Romani, Paubert, \&
  Encrenaz}]{Rosenqvist1992}
Rosenqvist, J., Lellouch, E., Romani, P., Paubert, G., \& Encrenaz, T. 1992,
  ApJL, 392, L99

\bibitem[{Teanby {et~al.}(2006)Teanby, Fletcher, Irwin, Fouchet, \&
  Orton}]{Teanby2006}
Teanby, N., Fletcher, L.~N., Irwin, P. G.~J., Fouchet, T., \& Orton, G.~S.
  2006, Icarus, 185, 466.
\newblock \url{http://linkinghub.elsevier.com/retrieve/pii/S0019103506002569}

\bibitem[{Thelen {et~al.}(2019)Thelen, Nixon, Chanover, Cordiner, Molter,
  Teanby, Irwin, Serigano, \& Charnley}]{Thelen2019}
Thelen, A.~E., Nixon, C., Chanover, N., {et~al.} 2019, Icarus, 319, 417.
\newblock \url{https://linkinghub.elsevier.com/retrieve/pii/S0019103518304184}

\bibitem[{Tollefson {et~al.}(2019)Tollefson, de~Pater, Luszcz-Cook, \&
  DeBoer}]{Tollefson2019}
Tollefson, J., de~Pater, I., Luszcz-Cook, S., \& DeBoer, D. 2019, The
  Astronomical Journal, 157, 251.
\newblock \url{http://dx.doi.org/10.3847/1538-3881/ab1fdf}

\bibitem[{Villanueva {et~al.}(2018)Villanueva, Smith, Protopapa, Faggi, \&
  Mandell}]{Villanueva2018}
Villanueva, G.~L., Smith, M.~D., Protopapa, S., Faggi, S., \& Mandell, A.~M.
  2018, Journal of Quantitative Spectroscopy and Radiative Transfer, 217, 86.
\newblock \url{https://doi.org/10.1016/j.jqsrt.2018.05.023}

\bibitem[{Vinatier {et~al.}(2015)Vinatier, B{\'{e}}zard, Lebonnois, Teanby,
  Achterberg, Gorius, Mamoutkine, Guandique, Jolly, Jennings, \&
  Flasar}]{Vinatier2015}
Vinatier, S., B{\'{e}}zard, B., Lebonnois, S., {et~al.} 2015, Icarus, 250, 95.
\newblock \url{http://dx.doi.org/10.1016/j.icarus.2014.11.019}

\bibitem[{Wayne(2000)}]{Wayne2000}
Wayne, R. 2000, {Chemistry of Atmospheres} (Oxford University Press)

\end{thebibliography}

\end{document}